\documentclass[aps,prb,groupedaddress]{revtex4-1}
\usepackage{dcolumn}
\usepackage{bm}
\usepackage[colorlinks,hyperindex, urlcolor=blue, linkcolor=blue,citecolor=black, linkbordercolor={.7 .8 .8}]{hyperref}
\usepackage{graphicx}
\usepackage{tabularx}
\usepackage{amsfonts}
\usepackage{amsmath}
\usepackage{amssymb}
\usepackage{amsbsy}
\usepackage{setspace}
\usepackage{nicefrac}
\usepackage{xfrac}
\usepackage{bm}
\usepackage{float}

\begin{document}

\newcommand*\arucl{{$\alpha$-RuCl$_3$}}

\newcommand*\rucl{{RuCl$_3$}}

\newcommand{\enquote}[1]{``#1''}

\title{Evidence for charge transfer and proximate magnetism in graphene/$\alpha$-RuCl$_3$ heterostructures}

\author{Boyi Zhou}
\author{J.~Balgley}
\affiliation{Department of Physics, Washington University in St.~Louis, 1 Brookings Dr., St.~Louis MO 63130, USA}
\author{P.~Lampen-Kelley}
\author{J.-Q.~Yan}
\author{D.~G.~Mandrus}
\affiliation{Material Science \& Technology Division, Oak Ridge National Laboratory, Oak Ridge, Tennessee 37831, USA}
\affiliation{Department of Material Science and Engineering, University of Tennessee, Knoxville, Tennessee 37996, USA}
\author{E.~A.~Henriksen}
\email{henriksen@wustl.edu}
\affiliation{Department of Physics, Washington University in St.~Louis, 1 Brookings Dr., St.~Louis MO 63130, USA}
\affiliation{Institute for Materials Science \& Engineering, Washington University in St.~Louis, 1 Brookings Dr., St.~Louis MO 63130, USA}

\date{\today}

\begin{abstract}
We report a study of electronic transport in van der Waals heterostructures composed of flakes of the antiferromagnetic Mott insulator \arucl~placed on top of monolayer graphene Hall bars. While the zero-field transport shows a strong resemblance to that of isolated graphene, we find a consistently $p$-type Hall effect suggestive of multiband conduction, along with a non-monotonic and gate-voltage-dependent excursion of the resistivity at low temperatures that is reminiscent of transport in the presence of a magnetic phase transition. We interpret these data as evidence for charge transfer from graphene to \arucl~in an inhomogeneous device yielding both highly- and lightly-doped regions of graphene, while the latter shows a particular sensitivity to magnetism in the \arucl. Thus proximity to graphene is a means to access magnetic properties of thin layers of magnetic insulators.
\end{abstract}

\maketitle

\section{Introduction}

The layered Mott insulator \arucl~exhibits phenomena consistent with quantum spin liquid behavior~\cite{kitaev_anyons_2006,sandilands_scattering_2015,banerjee_proximate_2016,baek_evidence_2017,do_majorana_2017-1,zheng_gapless_2017,banerjee_neutron_2017,janvsa_observation_2018}. Particularly intriguing among recent discoveries is a half-integer quantized thermal Hall conductance~\cite{kasahara_majorana_2018}, which may signal the presence of non-Abelian excitations useful in creating a topological quantum bit~\cite{kitaev_fault-tolerant_2003}. Recent studies of \arucl~employ a variety of bulk magnetic probes on high quality samples, mm- to cm-scale in size, which are generally found to behave as Kitaev paramagnets at temperatures above $T_{N\acute{e}el}$ of a zigzag antiferromagnet~\cite{sandilands_scattering_2015,baek_evidence_2017,do_majorana_2017-1,zheng_gapless_2017,banerjee_neutron_2017,janvsa_observation_2018}. Despite the convenience of electronic transport, it is not widely used due to the Mott insulating nature of \arucl~\cite{binotto_optical_1971,rojas_hall_1983,mashhadi_electrical_2018,noauthor_notitle_nodate,plumb_$alpha$-rucl$_3$_2014}. 

Seeking to probe \arucl~by electronic methods, we have studied the electronic transport in heterostructures comprised of \arucl~stacked on monolayer graphene. Incorporating graphene into stacks of various layered materials or thin films is a promising approach to discover new physics and potential applications~\cite{geim_van_2013}. In particular, graphene layered with various magnetic insulators including YIG, EuO, and EuS has been proposed as a platform for new magnetic phases or proximity-induced magnetism~\cite{haugen_spin_2008,yang_proximity_2013,zhang_proximity_2014,wang_proximity-induced_2015,wei_strong_2016,su_effect_2017}. Proximity effects for graphene in contact with the antiferromagnets BiFeO$_3$ and RbMnCl$_3$ have also been theoretically considered~\cite{qiao_quantum_2014,zhang_quantum_2015}, though in both cases the graphene interacts with ferromagnetically-aligned spins. Meanwhile, the precise nature of the interface of graphene with other materials is of much current interest, for instance in graphene layered with transition metal dichalcogenides where a charge transfer or even spin-orbit-proximity effect has been found~\cite{he_electron_2014,yuan_photocarrier_2018,lin_electron_2019,island_spin-orbit-driven_2019}. 

Here we explore transport in graphene next to an antiferromagnetic Mott insulator with the potential for quantum spin liquid physics~\cite{sandilands_scattering_2015,banerjee_proximate_2016,baek_evidence_2017,do_majorana_2017-1,zheng_gapless_2017,banerjee_neutron_2017,janvsa_observation_2018,kasahara_majorana_2018}. We find the transport in these devices---which at first glance is very similar to that of standard graphene-on-SiO$_2$---nonetheless shows a strongly enhanced conductivity and clear signatures of multi-band transport. While the presence of a Dirac peak associated with graphene would appear to suggest that no charge transfer has occurred despite the different work functions of graphene and \arucl, the Hall effect shows robust evidence for a sizable population of holes in coexistence with the standard graphene gate-voltage-dependent transport. Moreover between 15 and 40 K, we find a non-monotonic temperature dependence of the resistivity, suggestive of transport in the presence of magnetic phase transitions. While \arucl~has a $T_{N\acute{e}el}$ of 7 or 14 K depending on structural disorder~\cite{cao_low-temperature_2016}, here the implied critical temperatures are roughly twice as large. Altogether, our data are consistent with a picture of inhomogeneous transport where the graphene and \arucl~are intermittently in contact yielding regions that are either lightly- or highly-doped, the latter arising from an expected charge transfer between graphene and \arucl~\cite{gerber_ab_2019} that is also predicted to reinforce the antiferromagnetism. Finally, the gate-voltage dependence of the magnetic signatures suggests it occurs in the lightly-doped regions that in our picture are not in direct contact with the \arucl.

\section{Experiment}

\begin{figure*}[t]
\includegraphics[width=0.7\textwidth]{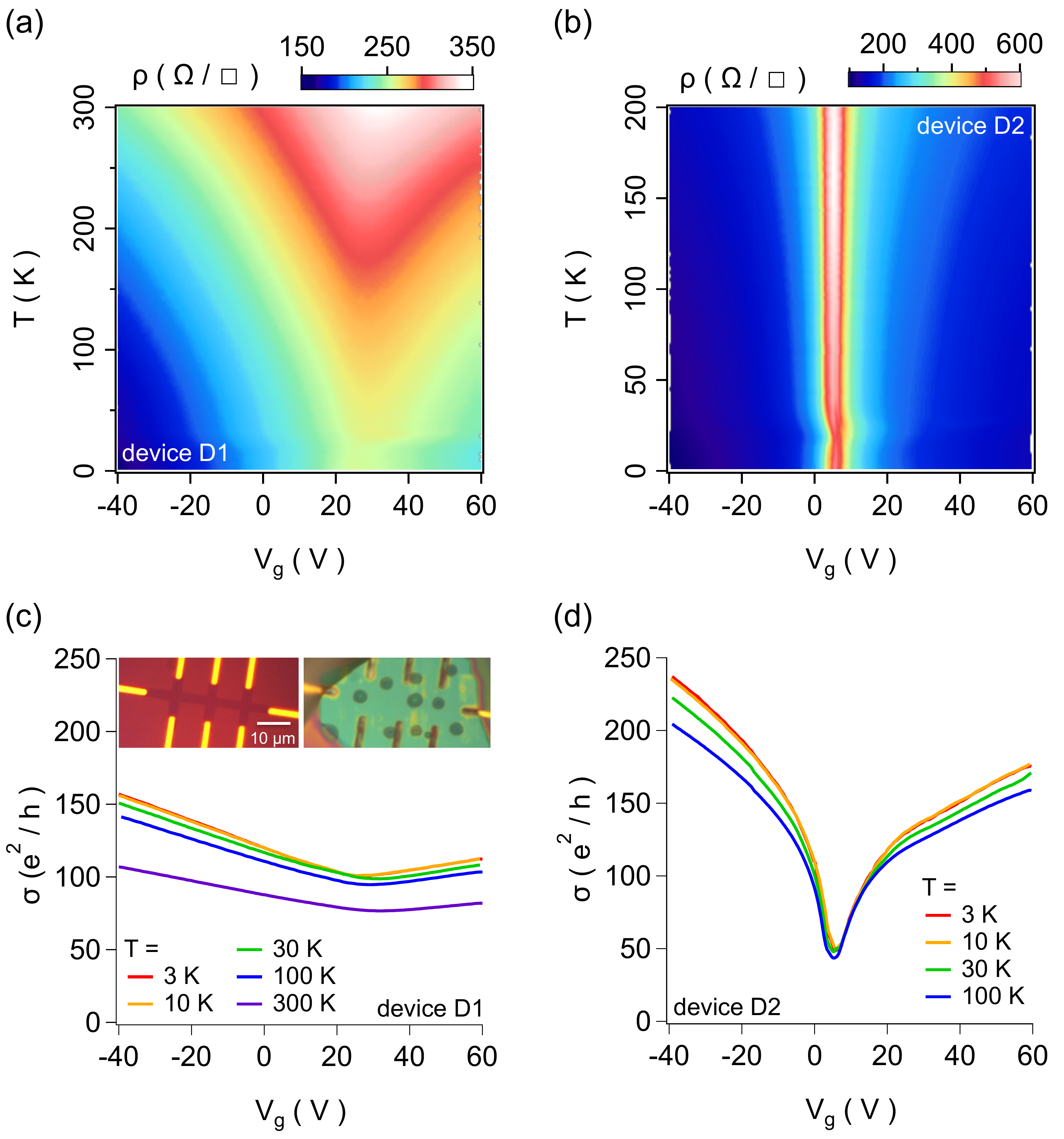}
\caption{\textbf{Transport at zero magnetic field.} Though similar in appearance to the usual electronic transport in graphene-on-oxide, the minimum conductivity is much larger than expected. \textbf{a-b}, Resistivity vs temperature and gate voltage in two representative devices, D1 and D2, respectively. \textbf{c-d}, Constant-temperature linecuts of \textbf{a} and \textbf{b}, replotted as the conductivity, $\sigma = 1/\rho$. Inset to \textbf{c} are images of device D1 showing the monolayer graphene Hall bar before (left) and after (right) being covered by a ${\sim}10$-nm-thick \arucl~flake. Circular features are remnant spots from a polycarbonate layer used in transferring the flake.  \label{cond}} 
\end{figure*}

Single crystals of \arucl~were grown using a vapor transport technique from phase pure commercial \arucl~powder~\cite{banerjee_neutron_2017}. The devices consist of monolayer graphene exfoliated on Si wafers with a 300-nm-thick surface oxide layer. The graphene is etched into a Hall bar pattern using a patterned polymethyl-methacrylate mask and an O$_2$ plasma, followed by standard thin film patterning for contacts made of 3/30 nm of Cr/Au. The graphene surface is then cleaned by sweeping with an atomic force microscope tip in contact mode, which serves to remove the remnant nm-thick layer of electron beam resist~\cite{goossens_mechanical_2012,lindvall_cleaning_2012,lin_graphene_2012,jalilian_scanning_2011,noauthor_notitle_nodate}. A flake of \arucl, exfoliated from parent crystals onto separate oxidized wafers~\cite{zhou_possible_2018}, is then transferred on top of the graphene using a polycarbonate film stretched over a small silicone stamp. The \arucl~flakes range in thickness from $5{-}25$ nm (${\sim}10{-}40$ layers) thick; prior Raman spectroscopy of flakes of comparable thickness give the same spectra as a pristine bulk sample~\cite{zhou_possible_2018}. Images of a typical device are shown inset to Fig.~\ref{cond}\textbf{c} before and after transferring the \arucl~flake. All measurements were performed using standard low-frequency lockin techniques in a variable temperature cryostat with a 9 T magnet, using gate voltages applied to the Si substrates. The graphene carrier density was determined either directly from analysis of Shubnikov-de Haas oscillations, or by known calibrations for the wafers used in these devices~\cite{elias_electronic_2017}, $n {=} 7.2 \times 10^{10} \times (V_g{-}V_{DP})$ cm$^{-2}$V$^{-1}$, where $V_{DP}$ is the voltage at which the Dirac peak is observed.

\section{Results}

\subsection{Zero-field transport}

The four-terminal resistivity at zero magnetic field of two representative devices is shown in Fig.~\ref{cond}\textbf{a} and \textbf{b} vs both the back gate voltage, $V_g$, and temperature, $T$. These data are clearly akin to typical graphene-on-oxide transport:~a maximum in the resistivity (``Dirac peak'') appears as $V_g$ is swept, which is rather broad in Fig.~\ref{cond}\textbf{a} and narrower in Fig.~\ref{cond}\textbf{b}. All devices explored show similar behavior~\cite{noauthor_notitle_nodate}, with a range of Dirac peak widths and gate voltage locations of the peak ($V_{DP}$). A decrease in the resistivity with temperature is seen that is consistent with a reduction in phonon scattering~\cite{chen_intrinsic_2008}.

\begin{figure*}[t]
\includegraphics[width=0.8\textwidth]{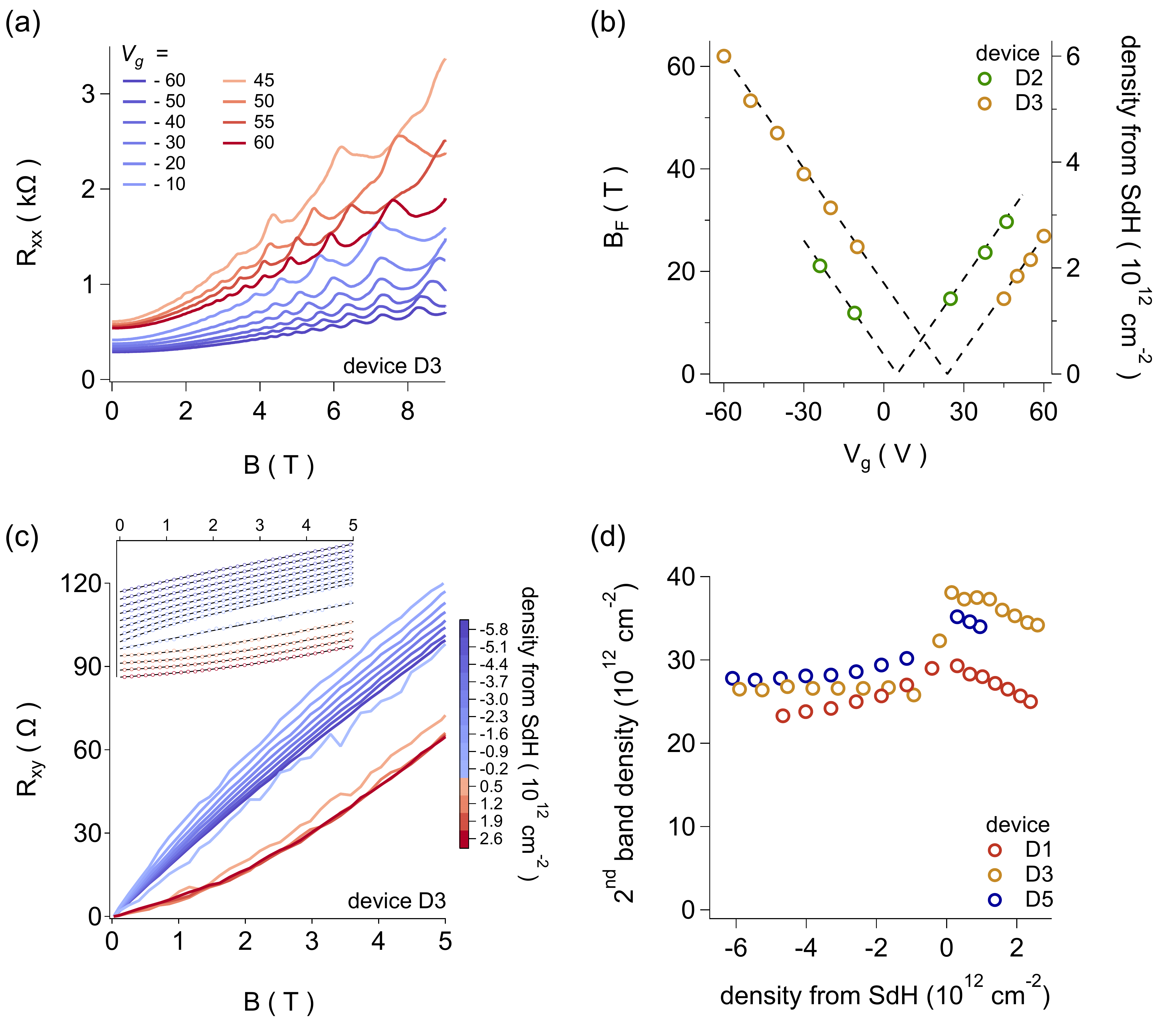}
\caption{\textbf{Magnetotransport of graphene/\arucl~devices.} \textbf{a} Shubnikov-de Haas oscillations from device D3 for a range of gate voltages on either side of the minimum conductivity (here at $V_g{=}25$ V). \textbf{b} The SdH oscillation frequency, $B_F$, for two devices. The corresponding charge density $n{=}g B_f/\phi_0$ (where we assume $g{=}4$ degrees of freedom as for graphene and $\phi_0$ is the magnetic flux quantum) is marked on the right axis. Dashed lines show the expected graphene density, $n{=}\alpha (V_g{-}V_{DP})$, based on prior measurements of isolated graphene flakes on the same substrates, with $\alpha{=}7.2\times10^{10}$ carriers/cm$^{2}$/V and $V_{DP}$ is the location of the zero-field conductivity minimum (see Methods). \textbf{c} Main panel:~Hall resistance for D3 acquired over a range of gate voltages (converted to carrier densities by the calibration in \textbf{b}). The colors correspond to those in \textbf{a}, and include three additional traces at densities close to charge neutrality that did not exhibit SdH oscillations. Inset:~the same data, vertically offset for clarity, with fits (black lines) to a two-band model of magnetotransport. \textbf{d} Carrier densities of the second band determined from two-band model fits.  \label{2band}} 
\end{figure*}

A marked departure from standard graphene transport becomes clear in Figure~\ref{cond}\textbf{c} and \textbf{d} where we show constant temperature profiles from Fig.~\ref{cond}\textbf{a} and \textbf{b}, respectively, re-plotted as the conductivity, $\sigma$. For all traces, $\sigma$ increases monotonically with increasing gate voltage to either side of the local conductivity minimum, $\sigma_{min}$, which in isolated graphene marks the charge neutrality point (here $\sigma_{min}$ occurs at $V_{DP}{=}{+}23$ V and $+5$ V for D1 and D2, respectively). However, the values of $\sigma_{min}$ in \emph{every} device are anomalously large, with values as high as 50 or 100 e$^2$/h at $T{=}3$ K; the highest we have found so far is 240 e$^2$/h. This stands in sharp contrast to the typical $\sigma_{min} {=} 2{-}12$ e$^2$/h routinely observed in regular graphene-on-oxide devices~\cite{tan_measurement_2007,adam_self-consistent_2007,chen_charged-impurity_2008}. To directly verify this  conductivity enhancement, we fabricated a long Hall bar and transferred flakes of \arucl~on one half and hexagonal boron nitride on the other, and found the resistivity of the \arucl-covered region to be an order of magnitude lower than the boron nitride covered portion~\cite{noauthor_notitle_nodate}.

\begin{figure*}[t]
\includegraphics[width=\textwidth]{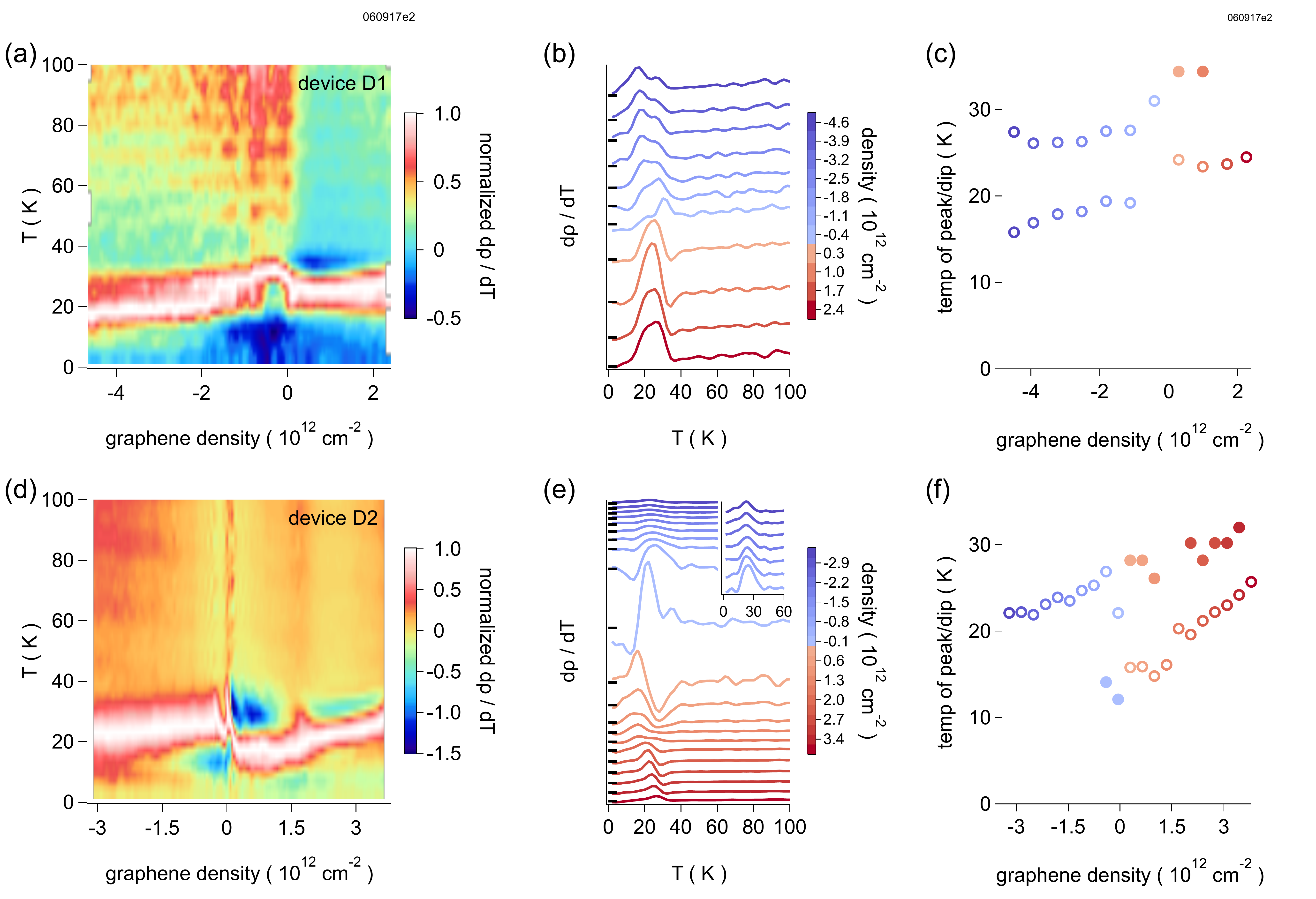}
\caption{\textbf{Transport signatures of magnetic transitions}. The resistivity of graphene/\arucl~heterostructures shows  characteristics of magnetic transitions~\cite{de_gennes_anomalies_1958,fisher_resistive_1968,alexander_critical_1976,rapp_electrical_1978,ausloos_critical_1980}. These data are for the same two devices as Fig.~\ref{cond}:~D1 (\textbf{a-c}) and D2 (\textbf{d-f}). \textbf{a,d} Colormaps showing $d\rho/dT$ vs the graphene charge carrier density in each device, normalized by the maximum value of $d\rho/dT$ below 60 K at each density. This highlights the evolution in temperature at which peaks in $d\rho/dT$ are observed to occur. \textbf{b,e} Linecuts of $d\rho/dT$ from \textbf{a} and \textbf{d}, offset vertically and not normalized in order to show the variation in amplitude of the peaks and dips. Blue-to-red shading follows the transition from $p$- to $n$-type doping of the graphene, with charge neutrality at the blue/red border. The inset to \textbf{e} shows a $12\times$ magnified view of the top $p$-type traces. Short black lines along the left axis mark the location of $d\rho/dT{=}0$ for each trace. \textbf{c,f} Temperatures of the $d\rho/dT$ peak maxima (open circles) and dip minima (filled circles). For D1 (\textbf{a-c}) two peaks can be discerned for $p$-type graphene. In both devices, the peak temperatures show a sharp variation at charge neutrality in graphene. \label{cr}} 
\end{figure*}

\subsection{Low-field transport}

Shubnikov-de Haas oscillations can be discerned in the magnetoresistance of some devices, as shown in Fig.~\ref{2band}\textbf{a} for a range of gate voltages on either side of the conductivity minimum for device D3. In Fig.~\ref{2band}\textbf{b} we plot the frequency, $B_F$, of the oscillations in $1/B$ as a function of $V_g$ for devices D2 and D3, and calculate the corresponding density of charge carriers participating in the SdH oscillations from $B_F{=}n \phi_0/g$ assuming $g{=}4$ which counts the usual spin and valley degeneracies in graphene, and $\phi_0{=}h/e$ is the magnetic flux quantum. The densities derived in this way correlate precisely with prior calibrations of the charge-gating efficiency of our Si/SiO$_2$ substrates. In other words, the SdH oscillations reveal bipolar behavior and a charge neutrality point, consistent with the usual graphene picture but for the enhanced conductivity which, in this device, appears as a minimum conductivity at $T{=}2$ K of 63 e$^2$/h.

In contrast, the Hall resistance $R_{xy}$ shown in Fig.~\ref{2band}\textbf{c} is markedly different from typical graphene behavior. In particular, (i) $R_{xy}$ is nonlinear, with (ii) values well below the Hall resistance of isolated graphene at equivalent charge densities, and (iii) surprisingly, as the graphene charge density is gated from $p$- to $n$-type, the Hall resistance does not change sign; indeed, the sign remains consistent with overall $p$-type transport. Since altogether we observe graphene-like behavior albeit with a higher conductivity and unusual Hall response, we analyze the data in a two-band model of graphene transport plus a second conducting band whose origin we will discuss later. In the inset to Fig.~\ref{2band}\textbf{c} we re-plot the Hall data with the traces offset for clarity, and perform curve fits using the standard two-band formalism~\cite{beer_galvanomagnetic_1963},
\begin{equation}
R_{xy} = \frac{B}{e}\frac{n_1 \mu_1^2 + n_2 \mu_2^2 + (n_1+n_2) (\mu_1 \mu_2 B)^2}{(|n_1|\mu_1 + |n_2| \mu_2)^2 + (n_1+n_2)^2 (\mu_1 \mu_2 B)^2}~,
\end{equation}
\noindent where $n_i$ and $\mu_i$ are the density and mobility of the $i^{th}$ band. For $n_1$ we use the graphene densities calibrated either from SdH oscillations or using the known gating efficiency for graphene-on-oxide, see Fig.~\ref{2band} and Methods. We also require the fitting coefficients to reproduce the zero-field conductivity, $\sigma_0 {=} e (n_1 \mu_1 + n_2 \mu_2)$. The resulting charge density of the second band, determined from measurements in three devices, is plotted in Fig.~\ref{2band}\textbf{d} vs.~the SdH-derived density. The sign of $n_2$ is hole-like, the magnitude is roughly an order of magnitude larger than the densities in the graphene, and there is a weak dependence on the density of the graphene band with either a step or peak at the charge neutrality point. The good fit to the Hall data and the consistent results among different samples is strong evidence for the presence of a second conducting band. Plots of the mobilities extracted for each band are included in the Supplemental Material~\cite{noauthor_notitle_nodate}. The graphene band mobilities are typically $2000{-}6000$ cm$^2$/Vs, reasonable for graphene-on-oxide devices; the mobilities of the second band lie between 500 and 2000 cm$^2$/Vs.  

\subsection{Temperature-dependent transport}

Returning to Fig.~\ref{cond}\textbf{a} and \textbf{b}, an additional feature is visible around 20 K that is not normally present in graphene. We highlight this by plotting the normalized temperature derivative of the resistivity for devices D1 and D2 in Fig.~\ref{cr}\textbf{a-c} and \textbf{d-f}, respectively (corresponding linecuts of the resistivity itself are included in the Supplemental Material). Intriguing lineshapes appear with peaks and dips whose specific shapes are distinctly different for $p$- or $n$-type graphene, and for which the peak temperatures show a sharp drop right at charge neutrality. We tentatively attribute this behavior to the presence of magnetic phase transitions. Bulk \arucl~is a zigzag antiferromagnet with $T_{N\acute{e}el}=7$ or 14 K depending on the stacking order~\cite{cao_low-temperature_2016}. It is well-known that the electrical resistivity can be impacted by magnetic transitions~\cite{de_gennes_anomalies_1958,fisher_resistive_1968}, with the shape of $d\rho/dT$ in the neighborhood of a magnetic transition being generically linked to the nature of the magnetism, e.g.~a peak is often associated with ferromagnetism where $T_{Curie}$ is at the peak maximum, and a peak/dip structure is expected for antiferromagnets with $T_{N\acute{e}el}$ at the dip minimum~\cite{alexander_critical_1976,rapp_electrical_1978,ausloos_critical_1980}. By analogy to this prior literature we suggest that a magnetic phase transition occurs in, or near, the graphene/\arucl~interface, and that the nature of this transition depends on the charge state of the graphene. While concrete identification of distinct phases is preliminary, the shape of $d\rho/dT$ is clearly correlated with the graphene charge carrier density and the visible Dirac peak; we include data for additional samples in the Supplemental Material~\cite{noauthor_notitle_nodate}. The transition temperatures implied by the peak and dip locations in Fig.~\ref{cr} lie in the range of $12{-}35$ K, rather higher than the 7 K or 14 K antiferromagnetic transition in bulk \arucl~\cite{cao_low-temperature_2016}. 

\section{Discussion}

In sum, in heterostructures of graphene next to \arucl~we observe a conductivity enhancement over that of isolated graphene, $p$-type Hall effect, and signatures of magnetic transitions. The differing work functions of graphene, at 4.6 eV~\cite{yu_tuning_2009}, and \arucl, at 6.1 eV~\cite{pollini_electronic_1996}, strongly imply a transfer of electrons from graphene to \arucl~will take place when the two materials come in contact, and indeed recent \emph{ab initio} calculations predict a charge transfer from monolayer graphene to monolayer \arucl~at the level of 4.7\% of an electron per Ru atom, along with a concomitant change in the magnetic couplings in \arucl~\cite{gerber_ab_2019}. However in the graphene/\arucl~heterostructure, (1) the gate voltage location of the Dirac peak---which is highly sensitive to extrinsic charge doping~\cite{chen_charged-impurity_2008,elias_electronic_2017}---is well within the usual range for graphene-on-oxide devices, suggesting no significant charge transfer from graphene has taken place, yet (2) the Hall effect shows clear evidence for two-band transport with population of holes well in excess of what is expected for the small shift of the Dirac peak away from $V_g{=}0$. We resolve this by proposing the coexistence of both lightly- and highly-doped regions of the graphene, due to the graphene and \arucl~flake not being in uniform contact. Wherever the two sheets are in close contact, the charge transfer expected from the work function difference occurs, but wherever they are separated the graphene remains nominally undoped. Indeed vdW devices often exhibit bubbles between the layers, and it is plausible the oxide-supported graphene inherits a nm-scale roughness so only higher-lying parts make contact with the overlying \arucl~flake. Thus transport measurements can yield a clear Dirac peak due to the intrinsic graphene, in parallel with highly-hole-doped graphene with a second charge neutrality point well outside the accessible $V_g$ range. In principle, since the \arucl~will have gained the balancing charge density, it may become conducting as well though this is not resolved here. We note that given the factor of 6 difference in areal density of C atoms in graphene to Ru atoms in \arucl, the predicted charge transfer~\cite{gerber_ab_2019} corresponds to a loss of roughly 0.8\% electrons/C from the graphene, or $2.8{\times}10^{13}$ cm$^{-2}$, which is remarkably close to the density of holes found in the second band in the nonlinear Hall analysis. Recently this charge transfer has been observed in graphene/\arucl~devices fabricated on boron nitride, but without the attendant magnetic signatures~\cite{mashhadi_spin-split_2019}.

The $V_g$-dependence of $d\rho/dT$ generically reflects the presence of the charge neutrality point in graphene. This  suggests the magnetic response is arising from the lightly-doped graphene regions that lie very close to but do not contact the \arucl~flake. We cannot rule out a magnetic response from the highly-doped regions as well, but expect the lightly-doped regions should be more sensitive to large fluctuations of the spin correlations in \arucl~near a magnetic transition. Thus the inhomogeneous nature of the samples fortuitously provides a window on magnetic effects arising at or near the interface. We note the elevated magnetic transition temperatures inferred above are consistent with an enhancement of magnetic couplings expected from the graphene-\arucl~charge transfer~\cite{gerber_ab_2019}.

We conclude with some general observations: (1) The Mott insulating nature of \arucl~may play a role here. The band gap of strongly correlated materials prepared as thin films on metals can be reduced over the bulk value due to screening of the Coulomb interaction, as observed for C$_{60}$ on silver~\cite{hesper_strongly_1997}. While the density of states in a thick silver film certainly is more effective at screening than a monolayer of graphene, there is still the question of how an added ${\sim}3{\times}10^{13}$ cm$^{-2}$ electrons in \arucl~will impact its electronic structure. (2) No dependence on the \arucl~flake thickness has been seen. Since the \arucl~layers are weakly bound together, it may be that only the one or two \arucl~layers closest to graphene are impacted by proximity. This could amplify the already considerable magnetic anisotropy~\cite{kubota_successive_2015,majumder_anisotropic_2015}, and so account for the enhanced magnetic transition temperatures we observe. (3) The detailed transport behavior is certain to depend on the nature of the graphene/\arucl~interface, in particular by the presence of remnant disorder (e.g.~water and incidental adsorbates accrued during fabrication) or the possibility of surface reconstruction effects~\cite{ziatdinov_atomic-scale_2016}. Notably, the Ru atoms are arranged in a honeycomb lattice as for C in graphene, with the two lattices close to a 2:5 commensurability so that mo\'ire physics may not be irrelevant~\cite{bistritzer_moire_2011-1}.

\section{Conclusion}

We have studied the electronic transport in monolayer graphene devices with a proximate \arucl~flake. The transport shows signatures of undoped graphene conducting in parallel to a large population of holes. We interpret this as transport at an inhomogeneous interface composed of both lightly- and highly-doped graphene, the latter arising from a significant charge transfer of electrons from graphene to \arucl. The resistivity at low temperatures shows signs of a magnetic phase transition that we interpret as a proximity effect appearing in the transport through the lightly-doped graphene regions.

\begin{acknowledgements}
We wish to thank A.~Banerjee, S.~Biswas, E.~Gerber, E.-A.~Kim, A.~MacDonald, S.~Nagler, R.~Valenti, and J.~van den Brink for informative discussions. We acknowledge support from the Institute of Materials Science and Engineering at Washington University in St.~Louis. EAH, BZ, and JB acknowledge support under NSF DMR-1810305. DGM and PLK acknowledge support from the Gordon and Betty Moore Foundation EPiQS Initiative through Grant GBMF4416.
\end{acknowledgements}

\bibliographystyle{nsf_erik}

\end{document}